\documentclass[11pt]{article}

\def\be{\begin{equation}}
\def\ee{\end{equation}}
\def\ba{\begin{eqnarray}}
\def\ea{\end{eqnarray}}
\def\la{\langle}
\def\ra{\rangle}
\def\a{\alpha}
\def\b{\beta}

\def\h{\hskip 1cm}

\def\lo{\longrightarrow}

\usepackage{graphicx}
\begin{document}
\begin{titlepage}
\vspace{4cm}
\begin{center}{\Large \bf Dynamics of entanglement of bosonic modes on symmetric graphs}\\
\vspace{1cm}F.
Ghahari,\footnote{email:ghaharikermani@mehr.sharif.edu}
\hspace{0.5cm} V. Karimipour \footnote{email:vahid@sharif.edu}
\hspace{0.5cm} R. Shahrokhshahi \footnote{email:shahrokhshahi@mehr.sharif.edu}\\
\vspace{1cm} Department of Physics, Sharif University of Technology,\\
P.O. Box 11365-9161,\\ Tehran, Iran
\end{center}
\vskip 3cm
\begin{abstract}
We investigate the dynamics of an initially disentangled Gaussian
state on a general finite symmetric graph. As concrete examples we
obtain properties of this dynamics on mean field graphs (also called
fully connected or complete graphs) of arbitrary sizes. In the same
way that chains can be used for transmitting entanglement by their
natural dynamics, these graphs can be used to store entanglement. We
also consider two kinds of regular polyhedron which show interesting
features of entanglement sharing.
\end{abstract}
\end{titlepage}
\section{Introduction}
One of the basic problems of quantum information processing is the
problem of entanglement transformation or more generally
manipulation of entanglement. For long distances one usually uses
photons through optical fibres or free air to transmit entanglement
. However for short distances other methods are being explored which
are based on using discrete sets of interacting quantum systems such
as spins \cite{bose1, bose2, bose3, bk, datta} or harmonic
oscillators \cite{plenio1}, which by their natural dynamics can
generate and transmit entanglement. In particular in \cite{plenio1},
one dimensional lattices of harmonic oscillators coupled by two
different types of Hamiltonians, were studied and various phenomena
were investigated with regard to entanglement generation and
transmission. Among other things it was shown that the largest
amount of entanglement between two oscillators is always obtained
when one places them at the two ends of an open chain. This
maximality was attributed to the fact that in this case the two
oscillators have fewer neighbors to which they can become entangled.
It was argued in \cite{plenio1} that besides linear arrays of
oscillators, other geometries, in principle any arrangement
corresponding to weighted graphs are worth of study, since they can
act as building blocks of more complicated networks. In
\cite{plenio1} itself two other geometries, namely a $Y$ shape
geometry which mimics a beam splitter and another geometry
corresponding to an interferometer were
studied.\\

In this article we want to extend these considerations in one
particular direction, namely we want to study compact and symmetric
geometries, i.e. symmetric graphs of finite size. The basic
motivation is that in contrast to the geometries considered in
\cite{ciracfrustration, ak} which were suitable for transmission of
entanglement, finite graphs are suitable for storing entanglement.
As any other resource, entanglement needs to be stored for use in
later suitable times and hence in any complicated network, building
blocks which can store entanglement, should be implemented. In the
simplest electrical analogy we may think of finite geometries as
capacitors and linear arrays of the type considered in
\cite{plenio1} as resistors or transmission lines.\\ However in
contrast to the static properties of entanglement, for which various
symmetric graphs have been considered \cite{ak}, for our purpose,
only one type of symmetric graph seems to be useful, namely the mean
field or a fully connected graph. The reason is the very simple
temporal behavior of entanglement on these graphs, compared with the
complicated behavior of arbitrary graphs. In fact a system of
harmonic oscillators on a mean field graphs has only two natural
frequencies which makes the resulting time development of
entanglement quite simple and easily controllable, while for other
symmetric graphs, this is not the case. If as in \cite{bose1} we are
to extract entanglement at an optimal time, then it is of utmost
importance that the dynamics of entanglement follows a simple and
not a complicated pattern. \\

For that reason we mostly consider mean field clusters of arbitrary
sizes and determine how an originally disentangled set of harmonic
oscillators positioned on the nodes of such a cluster, when coupled
to each other, develop a pairwise entanglement between themselves.
How this entanglement develops in time, what is its maximum value,
and how it depends on the size of the cluster. We stress that our
general setting is apt for analysis of any symmetric graph and we
indeed include two other graphs for observing some other
phenomena.\\

The structure of this paper is as follows: In section \ref{gauss} we
briefly review the Gaussian states and their entanglement
properties, especially we remind the closed formula for entanglement
of Formation (EoF) \cite{ciraceof} which in the context of symmetric
graphs is more suitable than negativity as a measure of
entanglement. In section \ref{dyn} we study the dynamics of a
Gaussian state on an arbitrary symmetric graph and obtain closed
formulas for the EoF between any two sites as a function of time.
This formula reduces the calculation of the EoF to the
diagonalization of the adjacency matrix of the graph.\\
In section \ref{2g} we study in detail the simplest graph consisting
of two nodes.  In section \ref{mean} we specialize to the mean field
graphs where our concrete results are reported in figures
(\ref{Ertime}, \ref{Erhighlow}, \ref{Ermaxmean(c)}) and table
(\ref{tablemean}).\\

Finally in section (\ref{sharing}) we consider two other symmetric
graphs for comparison and draw some conclusions about the sharing of
entanglement which challenge the arguments of \cite{plenio1} on this
issue.

\section{Preliminaries on Gaussian States}\label{gauss}
In this section we collect the rudimentary material on Gaussian
states that we need in the sequel. References \cite{tut, ferraro}
can be consulted for rather detailed reviews on the subject of
Gaussian states.\\ Let $\hat{R}:=(\hat{x}_1, \hat{x}_2, \cdots
\hat{x}_N, \hat{p}_1, \hat{p}_2, \cdots \hat{p}_N)$ be $N$ conjugate
operators characterizing $N$ modes and subject to the canonical
commutation relations
$$
    [\hat{R}_k,\hat{R}_l]=i\sigma_{kl},
$$ where $
    \sigma = \left(
\begin{array}{cc}
   & I_{n} \\
  -I_{n} &
\end{array}
    \right)
$
is the $2n$ dimensional symplectic matrix and $I_n$ denotes the $n$ dimensional unit matrix.\\
A quantum state $\rho$ is called Gaussian if its characteristic
function defined as
$
   C(\xi):= tr(e^{-i\xi_k\sigma_{kl} \hat{R}_l}\rho),
$
is a Gaussian function of the $\xi$ variables, namely if
$$
   C(\xi):= e^{\frac{-1}{2}\xi_k\Gamma_{kl}\xi_l},
$$
where we have assumed that linear terms have been removed by
suitable unitary transformations.  The matrix $\Gamma$, called the
covariance matrix of the state, encodes all the correlations in the
form
$$
\Gamma_{kl}:= \la R_kR_l+R_lR_k\ra - 2\la R_k\ra \la R_l\ra.
$$

For a two mode symmetric Gaussian state, one in which there is a
symmetry with respect to the interchange of the two modes, the
covariance matrix will be

\begin{equation}\label{gamma2}
    \Gamma = \left(\begin{array}{cc} \a & \b \\ \b & \a
    \end{array}\right).
\end{equation}
where the modes have been arranged in the order $x_1,\ p_1, \ x_2, \
p_2$ and $\a$ and $\b$ are $2\times 2$ symmetric matrices. By
symplectic transformations the covariance matrix of a two mode
symmetric Gaussian state  can always be put into the standard form
(in the order $x_1,x_2,p_1,p_2$)
\begin{equation}\label{f}
    \Gamma_s= \left(
    \begin{array}{cccc}
      n & k_x &  &  \\
     k_x  & n &  &  \\
       &  & n &k_p   \\
       & & k_p & n
    \end{array}
    \right),
\end{equation}
where $k_x\geq 0 \geq k_p$ and $k_x\geq |k_p|$. The entries of the
standard form of $\Gamma_s$ can be determined from the following
symplectic invariants:
\begin{eqnarray}
% \nonumber to remove numbering (before each equation)
  n^2 &=& \det \a  \\
  k_xk_p &=&\det \b  \\
  (n^2-k_x^2)(n^2-k_p^2) &=& \det \Gamma .
\end{eqnarray}
For a symmetric Gaussian state a closed formula for the entanglement
of Formation has been derived in \cite{ciraceof}. Note that there
are other criteria for studying the entanglement or separability of
Gaussian states \cite{pancini}, however we use only Entanglement of
Formation here to take advantage of the inherent built-in symmetry
of our graphs. EoF of a Gaussian state $\rho$, denoted simply by
$E(\rho)$ is expressed as follows:
\begin{equation}\label{g}
    E(\rho):= C_+\log_2 C_+-C_-\log_2 C_-,
\end{equation}
in which
\begin{equation}\label{h}
    C_{\pm} = \frac{(1\pm \Delta)^2}{4\Delta},
\end{equation}
and
\begin{equation}\label{i}
\Delta := min(1, \delta:=\sqrt{(n-k_x)(n+k_p)}).
\end{equation}
Thus a state is entangled only if $\delta\leq 1$.\\

Note that $\delta$ can be expressed in terms of the original
covariance matrix. To express it we denote
\begin{eqnarray}
% \nonumber to remove numbering (before each equation)
  u &:=& \det \a \\
  v &:=& \det \b \\
  w &:=& \det \Gamma
\end{eqnarray}
 and
\begin{equation}
\xi := u^2+v^2-w.
\end{equation}
Then a simple calculation gives
\begin{equation}\label{delta}
    \delta^2
     =u - v -
     \sqrt{\frac{\xi-\sqrt{\xi^2-4u^2v^2}}{2}}-\sqrt{\frac{\xi+\sqrt{\xi^2-4u^2v^2}}{2}}.
\end{equation}
In the following sections we use this equation for calculating the
entanglement of a Gaussian state which is initially disentangled and
evolves in time under a quadratic hamiltonian.
\section{Dynamics of entanglement of Gaussian states on symmetric graphs}\label{dyn}
Consider a symmetric graph, having $n$-nodes, corresponding to an
adjacency matrix $\textbf{A}$ and a system of bosonic modes
corresponding to the vertices of this graph interacting by a
quadratic Hamiltonian. In this paper we consider a Hamiltonian of
the form
\begin{equation}\label{H}
    H=\frac{1}{2}\sum_{i=1}^n p_i^2 + \frac{1}{2} \sum_{\la i,j\ra}
    x_i^2+x_j^2 + c(x_i-x_j)^2,
\end{equation}

where the sum runs over adjacent nodes and $c$ is a coupling
constant. This Hamiltonian describes a simple mass-spring system of
the form first studied by Plenio \cite{plenio1} in the context of
entanglement dynamics. The above Hamiltonian can be written in the
compact form

\begin{equation}\label{Hmatrix}
    H=\frac{1}{2}R^{\dagger} \left(\begin{array}{cc} V & 0 \\ 0 &
    T\end{array}\right)R
\end{equation}

where $V$ and $T$ (here equal to I) are the potential and the
kinetic matrices. We include the case of arbitrary $T$ (but
commuting with $V$) for generality, since some other Hamiltonians
like the one in \cite{ciracfrustration} can be expressed in this way.\\

The dynamics of $R$ is easily determined by solving the equations of
motion

\begin{equation}\label{eqmotion}
    \frac{dR}{dt}=-i[R,H]
\end{equation}
or using $[R_i, R_j]=i\sigma_{ij}$

\begin{equation}\label{dRdt}
    \frac{dR}{dt}= \sigma \hat{H} R = \left(\begin{array}{cc} 0 & T \\ - V &
    0\end{array}\right)R.
\end{equation}
The solution of the above equation is given by
\begin{equation}\label{Rt}
    R(t) = exp({\left(\begin{array}{cc} 0 & T \\ - V &
    0\end{array}\right)t})R(0).
\end{equation}
The explicit form of the evolution matrix is found by writing it as

\begin{equation}\label{Ut}
    U(t):=exp({\left(\begin{array}{cc} 0 & T \\ - V &
    0\end{array}\right)t})= e^{\frac{1}{2}\sigma_x \otimes (T-V) + \frac{i}{2}\sigma_y \otimes
    (T+V)}.
\end{equation}

In order to find the explicit form of the evolution matrix we use
the following\\

\textbf{Lemma:} For any two commuting matrices $A$ and $B$, the
following identity holds:

\begin{equation}\label{ide}
    e^{A\otimes \sigma_+ - B\otimes \sigma_-} = I\otimes \cos
    \sqrt{AB} +  \left(\sigma_+\otimes \sqrt{\frac{A}{B}} -
    \sigma_-\otimes \sqrt{\frac{B}{A}}\right)\sin \sqrt{AB}.
\end{equation}

This lemma is proved by a simple application of the identity
$e^{i\theta \hat{n}\cdot \vec{\sigma}} = \cos \theta + i\sin \theta
\hat{n}\cdot \vec{\sigma} $.

Using the above lemma, we find the final form of the evolution
matrix
\begin{eqnarray}\label{Utfinal}
    U(t) &=& I\otimes \cos
    \sqrt{VT}t  +  \left(\sqrt{\sigma_+\otimes \frac{T}{V}} -
    \sigma_-\otimes \sqrt{\frac{V}{T}}\right) \sin \sqrt{VT}\cr
    &=& \left(\begin{array}{cc} \cos \sqrt{VT}t & \sqrt{\frac{T}{V}}\sin \sqrt{VT} t\\
     -\sqrt{\frac{V}{T}}\sin \sqrt{VT}t & \cos \sqrt{VT} t\end{array}\right)
\end{eqnarray}

For the case we consider in this article the kinetic matrix $T$ is
identity ($T=I$) and so with the definition $W:=\sqrt{V}$, $U(t)$
simplifies to

\begin{equation}\label{UtT=1}
    U(t) = I\otimes \cos
    Wt  +  \left(\sigma_+\otimes W^{-1}-
    \sigma_-\otimes W\right)\sin Wt
    = \left(\begin{array}{cc} \cos Wt & W^{-1}\sin W t\\
     -W\sin Wt & \cos W t\end{array}\right).
\end{equation}

From the definition of the covariance matrix we find
\begin{equation}\label{gammat}
    \Gamma(t) = U(t)\Gamma(0) U^T(t).
\end{equation}
Let us consider the case where the initial state is a completely
uncorrelated state with $\Gamma(0) = I$.

The covariance matrix as a function of time will then be given by

\begin{equation}\label{gammat0}
    \Gamma(t) = U(t)U^T(t) =\left(\begin{array}{cc} \cos^2 Wt + W^{-2} \sin^2 Wt &
    (W^{-1}-W) \sin Wt \cos Wt \\
     (W^{-1}-W)\sin Wt\cos Wt &
    \cos^2 Wt + W^2 \sin Wt\end{array}\right).
\end{equation}
The covariance matrix between any two modes (sites of the graph) is
determined by extracting only the sub-matrix pertaining to those two
sites. For this we need the matrix which diagonalizes $V$. Let
$\Omega^{-1}W\Omega = W_D $ where $W_D $ is a diagonal matrices with
diagonal elements $\omega_i$. The matrix $\Omega$ is the matrix
which diagonalizes the adjacency matrix of
the graph.\\

\begin{equation}\label{gammatomega}
   \Gamma(t)=(\Omega\oplus \Omega)\left(\begin{array}{cc} \Gamma_{xx}^D& \Gamma_{xp}^D\\
   \Gamma_{xp}^D& \Gamma_{pp}^D\end{array}\right)(\Omega\oplus
   \Omega)^{T}
\end{equation}

where

\begin{eqnarray}\label{basic1}
% \nonumber to remove numbering (before each equation)
\Gamma_{xx}^D &=& diag\ \ (\cos^2 \omega_i t+\omega_i^{-2} \sin^2
\omega_i t) \cr &&\cr \Gamma_{pp}^D &=&   diag\ \ (\cos^2 \omega_i
t+\omega_i^{2}\sin^2 \omega_i t) \cr  &&\cr \Gamma_{xp}^D &=& diag\
\ ((\omega_i^{-1}-\omega_i)\sin \omega_i t \cos \omega_i t).
\end{eqnarray}
Then the covariance matrix between any two modes say modes $1$ and
$2$ will be the form

\begin{eqnarray}\label{basic2}
\Gamma_{x_1,x_1}  &=& \Gamma_{x_2,x_2} = \sum_{i=1}^n\Omega_{1i}^2
({\Gamma_{xx}^D})_i \cr \Gamma_{x_1,x_2}  &=& \Gamma_{x_2,x_1} =
\sum_{i=1}^n\Omega_{1i}\Omega_{2i} ({\Gamma_{xx}^D})_i \cr
\Gamma_{p_1,p_1}  &=& \Gamma_{p_2,p_2} = \sum_{i=1}^n{\Omega_{1i}}^2
({\Gamma_{pp}^D})_i \cr \Gamma_{p_1,p_2}  &=& \Gamma_{p_2,p_1} =
\sum_{i=1}^n\Omega_{1i}\Omega_{2i}({\Gamma_{pp}^D})_i \cr
\Gamma_{x_1,p_1}  &=& \Gamma_{x_2,p_2} = \sum_{i=1}^n\Omega_{1i}^2
({\Gamma_{xp}^D})_i \cr \Gamma_{x_1,p_2}  &=& \Gamma_{x_2,p_1} =
\sum_{i=1}^n\Omega_{1i}\Omega_{2i} ({\Gamma_{xp}^D})_i.
\end{eqnarray}

Therefore in each case we should only determine the matrix $\Omega $
which  diagonalizes the potential matrix $V$ and from (\ref{basic1},
and \ref{basic2} ) determine the eigenvalues $\omega_i$ .  In the
forthcoming sections we use this formalism to determine the dynamics
of entanglement between
any two modes on a wide variety of symmetric graphs. Note that the entanglement of formation is defined only for
symmetric Gaussian states, and in this paper we are considering only symmetric graphs. So in all of the graphs that we consider, this entanglement is invariant under
isomorphism of graphs.  \\

\section{The simplest example, A Two-Mode System}\label{2g}
As the simplest example we consider a two mode system represented by
a simple graph consisting of two nodes and a link connecting them.

The Hamiltonian is
\begin{equation}\label{H2}
    H=\frac{1}{2}(p_1^2+p_2^2)+\frac{1}{2}(x_1^2+x_2^2+c(x_1-x_2)^2),
\end{equation}
which corresponds to the matrices
\begin{equation}\label{V2}
T=I,\h \ \ and \ \  V=\left(\begin{array}{cc} 1+c & -c\\ -c &
1+c\end{array}\right).
\end{equation}
The eigenvalues of the matrix $V$ are readily obtained to be $1$ and
$\omega:=\sqrt{1+2c}$.

The covariance matrix is given by
\begin{equation}\label{Omega2}
    \Omega=\frac{1}{\sqrt{2}}\left(\begin{array}{cc} 1 & 1 \\ 1 &
    -1\end{array}\right),
\end{equation}
and
\begin{eqnarray}
% \nonumber to remove numbering (before each equation)
\Gamma^D_{x,x}&=&diag (1,\cos^2 \omega t + \omega^{-2} \sin^2 \omega
t)\cr \Gamma^D_{p,p}&=&diag (1,\cos^2 \omega t + \omega^{2} \sin^2
\omega t)\cr \Gamma^D_{x,p}&=&diag (0,(\omega^{-1}-\omega) \sin
\omega t \cos \omega
  t).
\end{eqnarray}
Using (\ref{delta}) we find the parameter $\delta$ which is
essential for calculating the entanglement of the state. The result
is
\begin{equation}\label{delta2}
    \delta =  \sqrt{1+\frac{1}{4}(\omega^{-1}-\omega)^2\sin^2 \omega t}-\frac{1}{2}
    \mid (\omega^{-1}-\omega)\sin \omega t \mid .
\end{equation}

\begin{figure}[t]
\centering
   \includegraphics[width=7cm,height=7cm,angle=0]{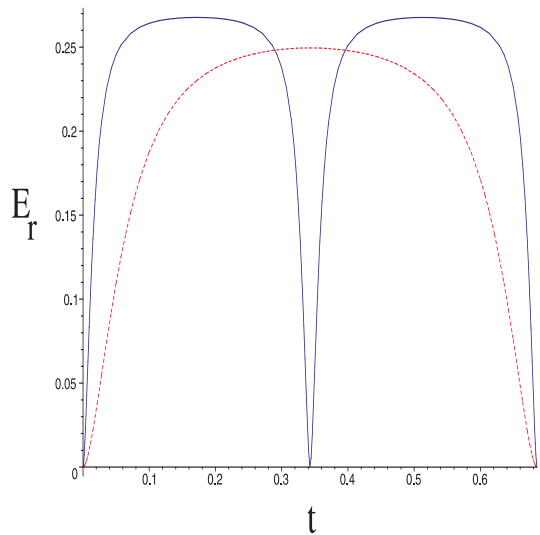}
   \caption{(color online) The entanglement between the two modes in a simple two-vertex graph, as a function of time,
   for two different couplings, $c=2$ (dashed line) and $c=8.3$ (solid line).
   } \label{2modesimple}
\end{figure}

The entanglement of the state is shown in figure (\ref{2modesimple})
for two difference coupling constants.

The entanglement oscillates at the natural frequency $\omega$. Its
maximum value is achieved for the minimum value of $\delta$, or
$$\delta_{min}=\delta(\omega
t=\frac{\pi}{2})=\sqrt{1+\frac{1}{4}(\omega^{-1}-\omega)^2}-\frac{1}{2}
    \mid (\omega^{-1}-\omega)\mid.$$
As we increase the frequency or the coupling constant, the
entanglement becomes flat in most of the period and develops cusp
singularities in half periods. Note that $\delta$ ranges between $0$
(for $\omega\lo \infty$) and $1$ (for $\omega=1$). Thus the maximum
entanglement increases unboundedly by increasing the strength of the
interaction $c$. In fact the maximum entanglement
increases as $\log \ \omega$ for large coupling constants $c$. \\
Thus a two-vertex graph can be used as a storage device for
entanglement the "capacity" of which increases logarithmically with
the coupling constant $c$. Moreover as the flatness of the curve in
figure (\ref{2modesimple}) shows, for very large coupling constants
we can extract this maximum entanglement at any time we wish except
for a discrete set
of points.\\

\section{Mean Field Clusters}\label{mean}
We now consider a mean field cluster of $N$ vertices in which every
vertex is connected to $N-1$ other vertices. The adjacency matrix
for a mean field graph is given by \be\label{a6} \hat{{\bf A}} =
E-I, \ee where $E$ is the matrix all of whose entries are equal to
1, $
    E_{ij}=1 \ \ \ \forall\ \ \  i,\ \ $  and $\  j $.
The potential matrix of this graph is given by
\begin{equation}\label{}
    \hat{V}=(1+Nc)I-cE.
\end{equation}

The matrix $E$ and hence ${\bf V}$ can easily be diagonalized. We
have
\begin{equation}
    Ee_0=Ne_0, \h E e_k=0, \ \ \ k=1,2,\cdots
    N-1,
\end{equation}
where
\begin{eqnarray}\label{a4}
% \nonumber to remove numbering (before each equation)
e_0 &=& \frac{1}{\sqrt{N}}(1,1,\cdots 1)^T, \cr e_k
&=&\frac{1}{\sqrt{k(k+1)}}(1,1,1,\cdots -k,\cdots 0)^T,\ \ \ \
k=1,2,\cdots N-1.
\end{eqnarray}

Thus the eigenvalues of ${\bf V}$ will be given by
\begin{equation}
% \nonumber to remove numbering (before each equation)
  \omega_0 =1 \h \omega_1=\cdots \omega_{N-1}=\sqrt{1+Nc}=:\omega
  .
\end{equation}

The eigenvectors $e_0$ to $e_{N-1}$ derived above easily yield the
diagonlizing matrix $\Omega$ ($\Omega_{ij}=(e_j)_i$) from which we
can obtain after straightforward calculations from (\ref{basic1})
and (\ref{basic2}) the following parameters of the covariance matrix
between any two sites say sites $1$ and $2$:

\begin{figure}[t]
\centering
   \includegraphics[width=8cm,height=8cm,angle=0]{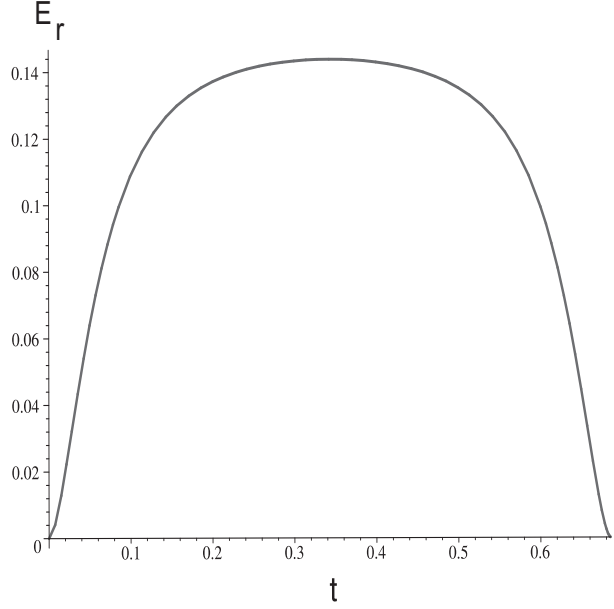}
   \caption{The re-scaled entanglement as a function of time for a mean field cluster of size $N=20$ at $c=1$, the time spans one full period.}\label{Ertime}
\end{figure}

\begin{eqnarray}\label{gammamean}
    \Gamma_{x_1,x_1}&=&\Gamma_{x_2,x_2}=\frac{1}{N}+(1-\frac{1}{N})(\cos^2
    \omega t+\omega^{-2} \sin^2 \omega t)\cr
   \Gamma_{x_1,x_2}&=&\Gamma_{x_2,x_1}=\frac{1}{N}(1-\omega^{-2}) \sin^2 \omega t\cr
 \Gamma_{p_1,p_1}&=&\Gamma_{p_2,p_2}=\frac{1}{N}+(1-\frac{1}{N})(\cos^2
    \omega t+\omega^{2} \sin^2 \omega t)\cr
 \Gamma_{p_1,p_2}&=&\Gamma_{p_2,p_1}=\frac{1}{N}(1-\omega^{2})\sin^2
    \omega t\cr
 \Gamma_{x_1,p_1}&=&\Gamma_{x_2,p_2}=(1-\frac{1}{N})(\omega^{-1}-\omega)\sin \omega t \cos\omega
 t\cr
 \Gamma_{x_1,p_2}&=&\Gamma_{x_2,p_1}=-\frac{1}{N}(\omega^{-1}-\omega)\sin \omega t \cos \omega t
    \end{eqnarray}

One can obtain the standard form of this matrix by using the
symplectic invariants. They read in the present case
\begin{eqnarray}\label{uvmean}
    u &=& \  1 + \frac{1}{N}(1-\frac{1}{N})(\omega-\omega^{-1})^2 \sin^2
    \omega t,\\
   v&=& \ \frac{-1}{N^2}(\omega-\omega^{-1})^2 \sin^2 \omega t,\\
w &=&\  1 + 2(\omega-\omega^{-1})^2 (\frac{N-2}{N^2})\sin^2 \omega
t.
\end{eqnarray}

\begin{figure}[t]
\centering
   \includegraphics[width=10cm,height=10cm,angle=0]{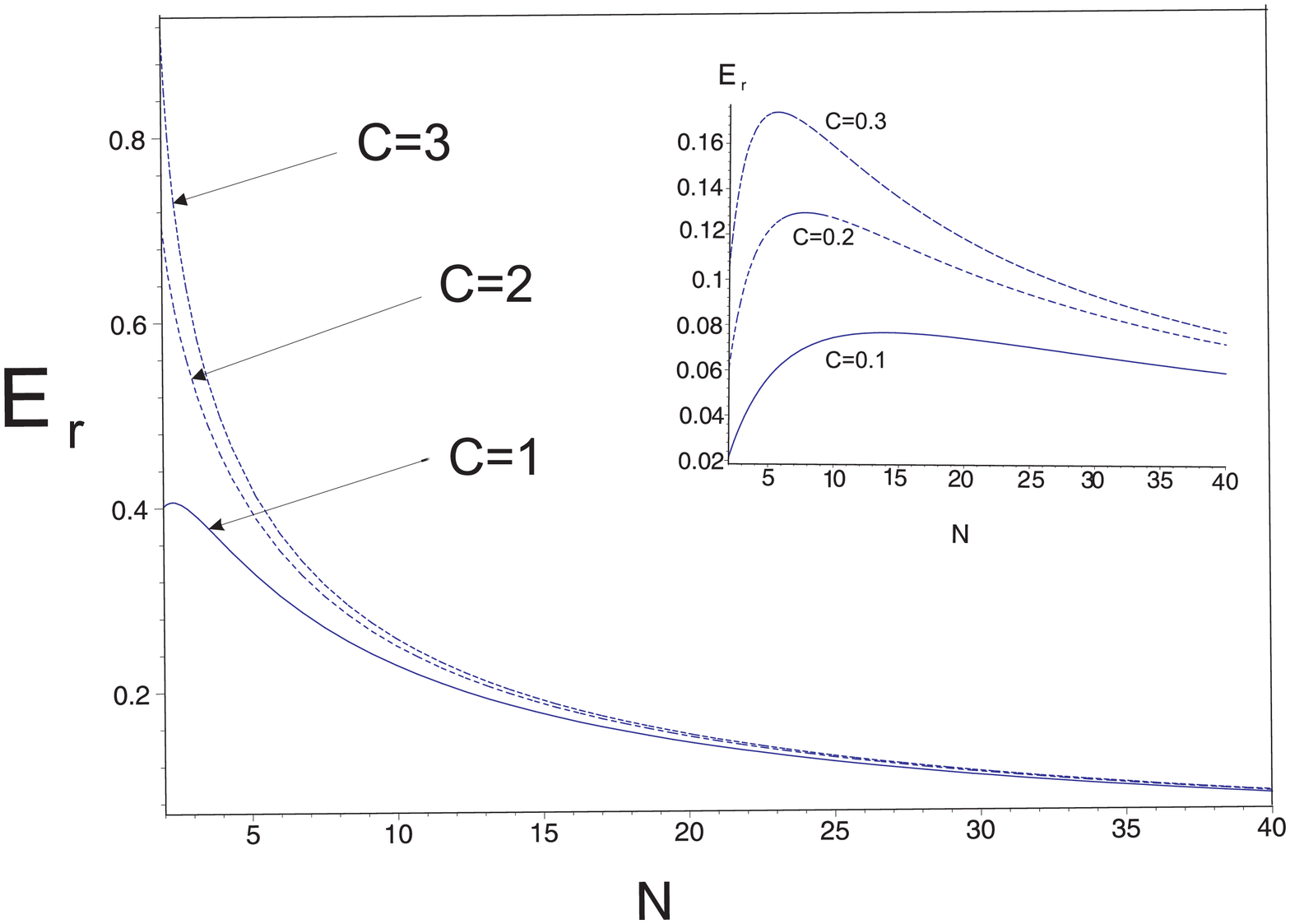}
   \caption{The maximum re-scaled EoF for mean field clusters for different coupling constants.} \label{Erhighlow}
\end{figure}

Inserting these values in \ref{delta} gives the entanglement for
these graphs. Following \cite{vidal}, we define re-scaled
entanglement $(E_r)$ which is $N-1$ times the entanglement between
any two nodes. This definition stems from the fact that a node
shares its entanglement with its neighbors which are $N-1$ in
number. \\
Figure (\ref{Ertime}) shows the re-scaled entanglement for a mean
field cluster of size 20 as a function of time. Figure
(\ref{Erhighlow}) show the maximum re-scaled entanglement for mean
field clusters as a function of their size for different coupling
constants.

\begin{figure}[t]
\centering
   \includegraphics[width=8cm,height=8cm,angle=0]{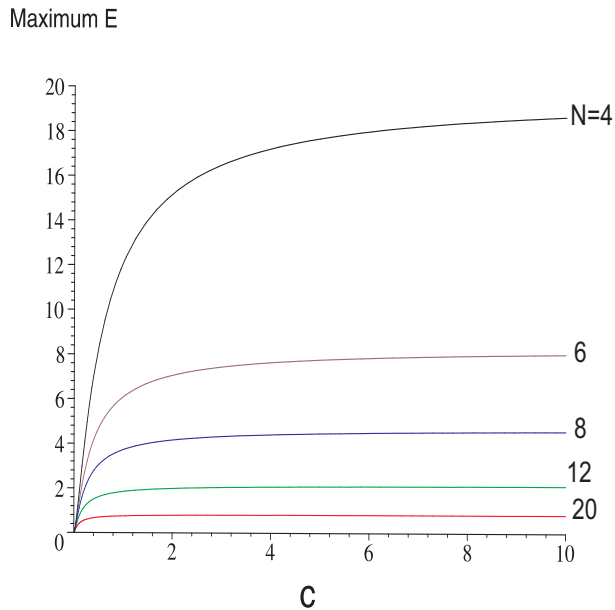}
      \caption{The amplitude of oscillation of entanglement ($E$) in units of 0.01 ebits, for mean field clusters of different
      sizes as a function of coupling constant. The saturation
      values are always less than the bounds found in \cite{ciracfrustration}.
      Note that for all the sizes $N>2$, as $c$ increases, $E$ saturates to finite values.} \label{Ermaxmean(c)}
\end{figure}

For any cluster of size $N$ and coupling constant $c$, the
entanglement oscillates at a frequency $\omega:=2\sqrt{1+Nc}$ and
most of the time the two modes have an appreciable amount of
entanglement. With increasing the coupling constant $c$, the
amplitude of oscillation increases and saturates to a finite value
for very large $c$, as long as $N>2$ (figure \ref{Ermaxmean(c)}).
Table (\ref{tablemean}) shows this saturated amplitude for clusters
of different sizes.
\begin{table}
\centering
\begin{tabular}{ccc}
\hline N & \vline & Maximum $E_r$ \\
\hline
2 & \vline & $\infty$\\
3 & \vline & $0.803$\\
4 & \vline & $0.592$\\
5 & \vline & $0.484$\\
6 & \vline & $0.415$\\
7 & \vline & $0.365$\\
8 & \vline & $0.328$\\
9 & \vline & $0.298$\\
10 & \vline & $0.274$\\
15 &\vline & $0.196$\\
20 &\vline & $0.156$\\
30&\vline& $0.113$\\
\end{tabular}
\caption{The saturated amplitude of re-scaled entanglement for mean
field graphs of different sizes.}
\end{table}\label{tablemean}

\section{Sharing of entanglement}\label{sharing}
\begin{figure}[t]
\centering
   \includegraphics[width=9cm,height=5cm,angle=0]{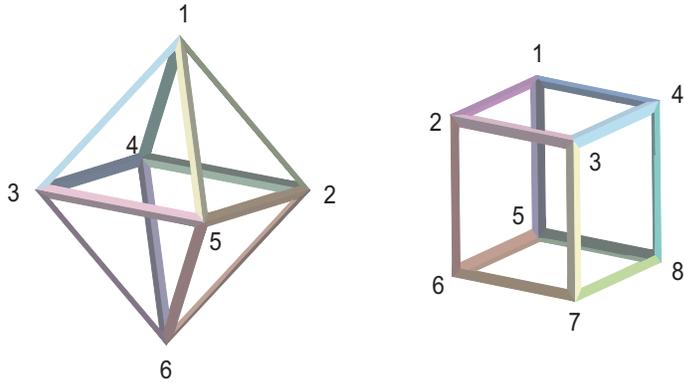}
      \caption{(color online) Two symmetric graphs, Cube and the Octahedron, the entanglement between different pairs of vertices
      are plotted in figure figure \ref{OctaH1} and \ref{CubeH1}.} \label{CubeOcta}
\end{figure}

As mentioned in the introduction, in \cite{plenio1} it was shown
that in a linear lattice, the largest amount of entanglement between
pairs of sites with the same distance, occurs when these two sites
are at the end points of the lattice.  This was attributed to the
fact that the endpoints of the lattice have fewer neighbors to which
they share their entanglement. In this regard it is instructive to
consider two symmetric graphs which are not fully connected. The two
graphs which we study are a six-vertex graph in the shape of
octahedron and an eight-vertex graph in the shape of a cube. They
are shown in figure (\ref{CubeOcta}) with numbered vertices.

\begin{figure}[t]
\centering
   \includegraphics[width=12cm,height=7.5cm,angle=0]{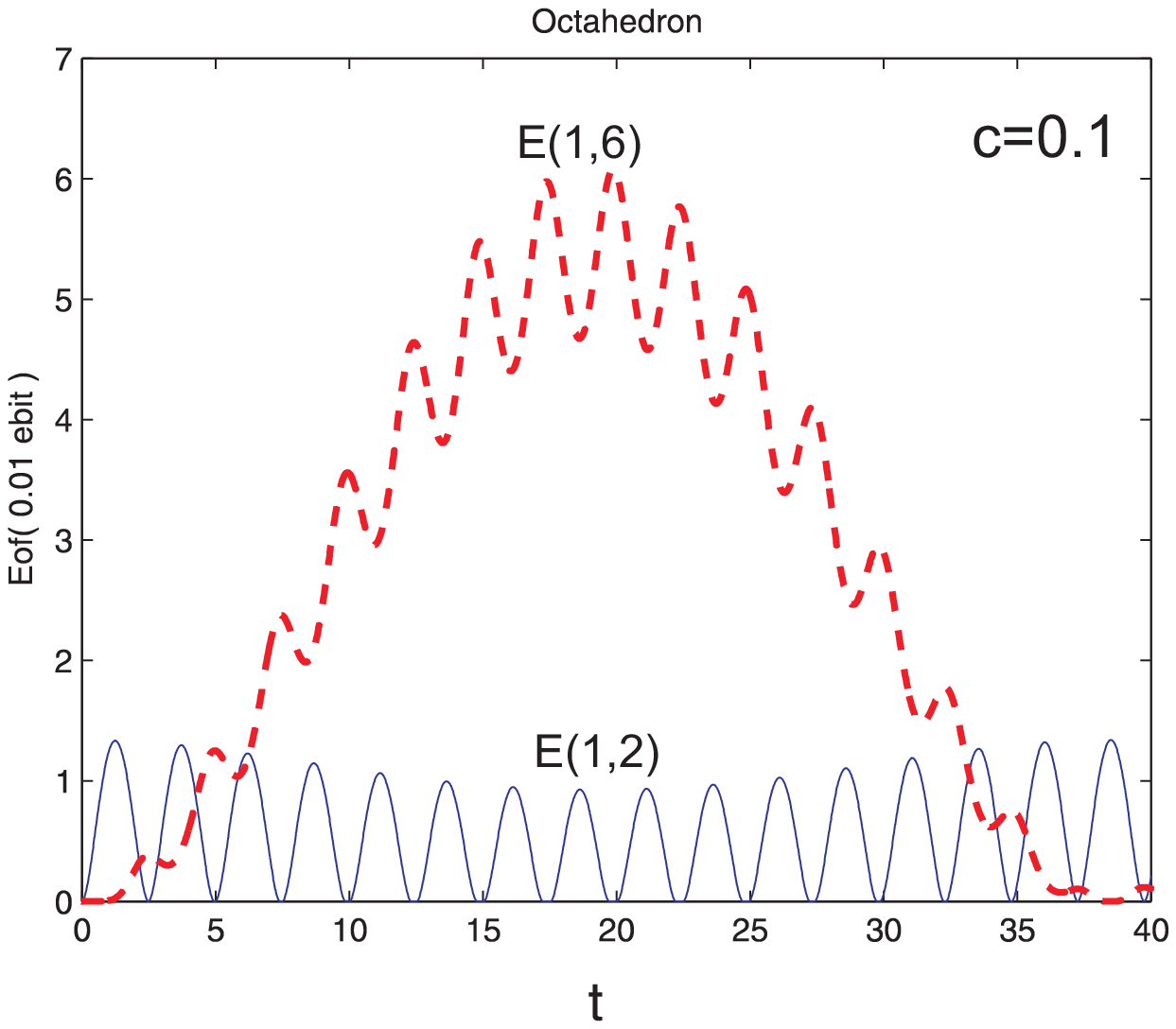}
      \caption{(color online) The pairwise entanglement between the sites $(1,2)$ and the sites $(1,6)$ in the
       Octahedron as a function of time.} \label{OctaH1}
\end{figure}

\begin{figure}[t]
\centering
   \includegraphics[width=12cm,height=7.5cm,angle=0]{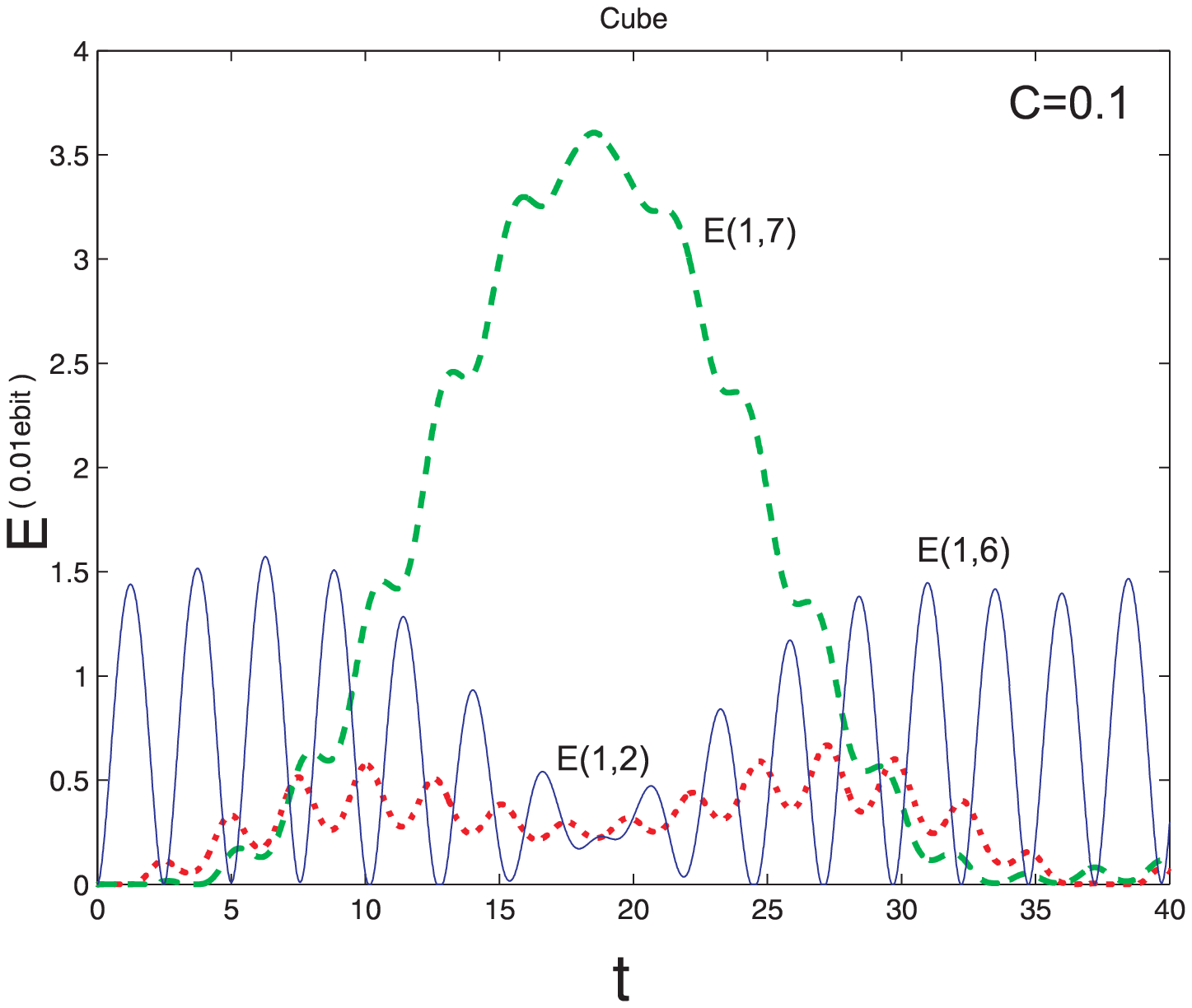}
      \caption{(color online) The pairwise entanglement between the sites $(1,2)$, $(1,3) $ and $(1,6)$ in the
       Cube as a function of time.} \label{CubeH1}
\end{figure}

In the octahedron there are essentially two types of pairs,
represented by the pair (1,2) and the pair (1,6). In each pair the
number of neighbors of each node is the same. However the pair (1,6)
although more apart than the pair (1,2) develops a much higher
entanglement, figure \ref{OctaH1}.
In terms of the number of edges, the distance between the nodes 1
and 2 is one, and there is only one shortest path which connects
these two nodes, while the distance between nodes 1 and 6 is two,
however there are four such shortest paths which connect these two
nodes. Therefore it seems that entanglement between two site is not
only affected by the number of their neighbors, but also by the
number of shortest paths which connects these two sites to each
other. To test this idea, we study the cube, which has three types
of pairs represented by the (1,2), (1,6) and (1,7), with distances
respectively given by 1, 2 and 3 and the number of shortest paths
respectively given by 1, 2, and 6. The entanglement is shown in
figure (\ref{CubeH1}) which confirms our assertion. Here we see a
competition between the two factors.
If we interpret the entanglement as a direct measure of quantum
correlations then these two figures show a very intriguing property
of entanglement: there are times where remote sites are strongly
quantum correlated while the nearest sites have a small quantum
correlation.

\section{Acknowledgements}
We would like to thank the members of the Quantum information group
of Sharif University for very valuable comments.

{}
\end{document}